\documentclass[aps,prl,twocolumn,groupedaddress,showpacs]{revtex4}
\usepackage{epsfig}

\def\bea {\begin{eqnarray}}
\def\eea {\end{eqnarray}}
\def\be {\begin{equation}}
\def\ee {\end{equation}}

\def\etal{{\it et al.}}
\def\F{{\cal F}}
\def\prl {Phys. Rev. Lett.\ }

\def\pr {Phys. Rev.\ }
\def\np {Nucl. Phys.\ }
\def\GV{G_{\mbox{\tiny V}}}
\def\GF{G_{\mbox{\tiny F}}}
\def\DRV{\Delta_{\mbox{\tiny R}}^{\mbox{\tiny V}}}
\def\fS{f_{\mbox{\tiny S}}}
\def\hyphen{{\mbox{-}}}

\def\2p{|2p\rangle }
\def\4p2h{|4p\hyphen 2h\rangle }
\def\6p4h{|6p\hyphen 4h\rangle }

\begin{document} 
\preprint{ }
 
\title{New limits on fundamental weak-interaction parameters from superallowed $\beta$ decay}

\author{J.C. Hardy}
\author{I.S. Towner}
\altaffiliation{Present address: 
Department of Physics,
Queen's University, Kingston, Ontario K7L 3N6, Canada}
\affiliation{Cyclotron Institute, Texas A\&M University,                    
College Station, Texas  77843}
\date{\today} 
\begin{abstract} 

A new critical survey of all world data on superallowed $0^{+} \rightarrow 0^{+}$ $\beta$ decays
provides demanding tests of, and tight constraints on, the weak interaction.  In confirmation of
the conserved vector current (CVC) hypothesis, the vector coupling constant, $\GV$, is demonstrated
to be constant to better than 3 parts in $10^4$, and any induced scalar current is limited to
$\fS \leq 0.0013$ in electron rest-mass units.  The possible existence of a fundamental scalar
current is similarly limited to $|C_S/C_V| \leq 0.0013$.  The superallowed data also determine
the up-down element of the Cabibbo-Kobayashi-Maskawa (CKM) matrix to be $V_{ud} = 0.9738(4)$.
With Particle Data Group values for $V_{us}$ and $V_{ub}$, the top-row test of CKM unitarity yields
$|V_{ud}|^2 + |V_{us}|^2 + |V_{ub}|^2 = 0.9966(14)$; although, if a recent result on $K_{e3}$ decay
is used exclusively to obtain $V_{us}$, this sum becomes 0.9999(16).  Either unitarity result can
be expressed in terms of the possible existence of right-hand currents.

\end{abstract} 

\pacs{23.40.Bw, 12.15.Hh, 12.60.-i}

\maketitle

Beta decay between nuclear analog states of spin-parity, $J^{\pi} = 0^+$, and isospin,
$T = 1$, has been a subject of continuous and often intense study for five decades.  The
strengths, or $ft$-values, of such transitions are nearly independent of nuclear-structure
ambiguities and depend uniquely on the vector part of the weak interaction.  Thus, their
measurement has given nuclear physicists access to clean tests of some of the fundamental
precepts of weak-interaction theory and, over the years, this strong motivation has led to
very high precision being achieved in both the experiments and the theory required to
interpret them.

As befits such an important issue, there have been periodic surveys of the relevant world
data (see, for example, Refs.~\cite{To73,Ha75,Ko84,Ha90}).  Because the last survey appeared
in 1990 and a large amount of new data has appeared in the decade and a half since then, we
have just completed a thorough new overview \cite{Ha05} in which we critically surveyed all relevant
measurements, adjusted original data to take account of the most modern calibration standards,
obtained statistically rigorous average results for each transition, and used updated and consistent
calculations to extract weak-interaction parameters from those results.  The outcome includes
more exacting confirmation of the conserved vector current (CVC) hypothesis, a reduced limit
on any possible scalar currents, and an improved value for the up-down quark-mixing element
of the Cabibbo-Kobayashi-Maskawa (CKM) matrix, $V_{ud}$.  The latter is an important component
of the most demanding available test of the unitarity of that matrix and can help constrain or
rule out the influence of right-hand currents.

For any superallowed $0^{+} \rightarrow 0^{+}$ $\beta$ decay between $T=1$ analog states, the
experimental $ft$-value can be related to the vector coupling constant {\it via} an expression
that includes several small ($\sim$1\%) correction terms.  It is convenient to combine some
of these terms with the $ft$-value and define a ``corrected" $\F t$-value.  Thus, we write \cite{To03}
\be
\F t \equiv ft (1 + \delta_R^{\prime}) (1 + \delta_{NS} - \delta_C ) = \frac{K}{2 \GV^2 
(1 + \DRV )}~,
\label{Ftconst}
\ee
where $K/(\hbar c )^6 = 2 \pi^3 \hbar \ln 2 / (m_e c^2)^5 = 8120.271(12) \times
10^{-10}$ GeV$^{-4}$s, $\GV $ is the vector coupling constant for semi-leptonic weak interactions,
$\delta_C$ is the isospin-symmetry-breaking correction and $\DRV$ is the transition-independent part
of the radiative correction.  The terms $\delta_R^{\prime}$ and $\delta_{NS}$ comprise the
transition-dependent part of the radiative correction, the former being a function only of the
electron's energy and the $Z$ of the daughter nucleus, while the latter, like $\delta_C$, depends in
its evaluation on the details of nuclear structure.  From this equation, it can be seen that
each measured transition establishes an individual value for $\GV$ and, if the CVC assertion is
correct that $\GV$ is not renormalized in the nuclear medium, all such values -- and all the
$\F t$-values themselves -- should be identical within uncertainties, regardless of the specific
nuclei involved.

The $ft$-value that characterizes any $\beta$-transition depends on three measured quantities: the
total transition energy, $Q_{EC}$, the half-life, $t_{1/2}$, of the parent state and the branching ratio,
$R$, for the particular transition of interest.  The $Q_{EC}$-value is required to determine the statistical
rate function, $f$, while the half-life and branching ratio combine to yield the partial half-life, $t$.
In our treatment \cite{Ha05} of the data, we considered all measurements formally published before November
2004 and those we knew to be in an advanced state of preparation for publication by that date.  We scrutinized
all the original experimental reports in detail and, where possible, adjusted those $Q_{EC}$-values that were
based on outdated calibration standards.  If corrections to any measurements were evidently required but
insufficient information was provided to make them, the results were rejected.  Of the surviving results,
only those with (updated) uncertainties that are within a factor of ten of the most precise measurement
for each quantity were retained for averaging.  The statistical procedures we followed in analyzing the data
were those used by the Particle Data Group in their periodic reviews of particle properties (e.g. \cite{PDG04}).

The final average $ft$ values obtained from the survey \cite{Ha05}, together with applied correction terms
and the resulting $\F t$ values, are given in Table~\ref{table1}
for the twelve superallowed transitions that are now known to 0.4\% precision or better; all but four are
actually known to better than 0.1\%.  They cover a broad range of nuclear masses from $A=10$ to $A=74$.  
As anticipated by CVC (see Eq.~\ref{Ftconst}) the $\F t$ values are statistically consistant with one
another, yielding an average value $\overline{\F t} = 3072.7(8)$ s, with a corresponding chi-square
per degree of freedom of $\chi^2/\nu = 0.42$.  These results have many important outcomes.  We will deal
with each, individually identified. 
  
\begin{table}
\begin{center}
\caption{Final average $ft$ values, correction terms, and $\F t$ values \cite{Ha05} for the twelve best known
superallowed transitions. 
\label{table1} }
\vskip 1mm
\begin{ruledtabular}
\begin{tabular}{lllll}
Parent & \multicolumn{1}{c}{$ft$} & \multicolumn{1}{c}{$\delta_R^{\prime}$} & \multicolumn{1}{c}{$\delta_C - \delta_{NS}$}
 & \multicolumn{1}{c}{$\F t$}  \\
Nucleus  & \multicolumn{1}{c}{(s)} & \multicolumn{1}{c}{(\%)} & \multicolumn{1}{c}{(\%)} & \multicolumn{1}{c}{(s)}  \\[1mm]
\hline 
& &  \\[-3mm]
~ $^{10}$C   & 3039.5(47)   & 1.652(4)  & 0.540(39) & 3073.0(49)        \\ 
~ $^{14}$O   & 3043.3(19)   & 1.529(8)  & 0.570(56) & 3071.9(26)        \\   
~ $^{22}$Mg  & 3052.4(72)   & 1.446(17) & 0.505(24) & 3080.9(74)        \\ 
~ $^{26m}$Al & 3036.7(12)   & 1.458(20) & 0.261(24) & 3072.9(15)        \\ 
~ $^{34}$Cl  & 3050.5(47)   & 1.425(32) & 0.720(39) & 3071.7(19)        \\
~ $^{34}$Ar  & 3059(12)     & 1.394(35) & 0.825(44) & 3076(12)          \\ 
~ $^{38m}$K  & 3051.1(10)   & 1.423(39) & 0.720(47) & 3072.2(21)        \\ 
~ $^{42}$Sc  & 3046.0(15)   & 1.437(47) & 0.460(47) & 3075.6(25)        \\ 
~ $^{46}$V   & 3045.5(22)   & 1.429(54) & 0.465(33) & 3074.7(30)        \\ 
~ $^{50}$Mn  & 3044.5(15)   & 1.429(62) & 0.547(37) & 3071.1(27)        \\ 
~ $^{54}$Co  & 3047.4(15)   & 1.428(71) & 0.639(43) & 3071.2(28)        \\ 
~ $^{74}$Rb  & 3083.8(75)   & 1.49(12)  & 1.50(41)  & 3083(15)          \\
 & & & Average, $\overline{\F t}$  & 3072.7(8)   \\
 & & & ~~~~~~~~~~$\chi^2/\nu$ & \multicolumn{1}{c}{0.42}   \\
\end{tabular}
\end{ruledtabular}
\end{center}
\end{table}

\begin{figure}[b]
\vspace{0.0cm}
\hspace{0.0cm}
\epsfig{file=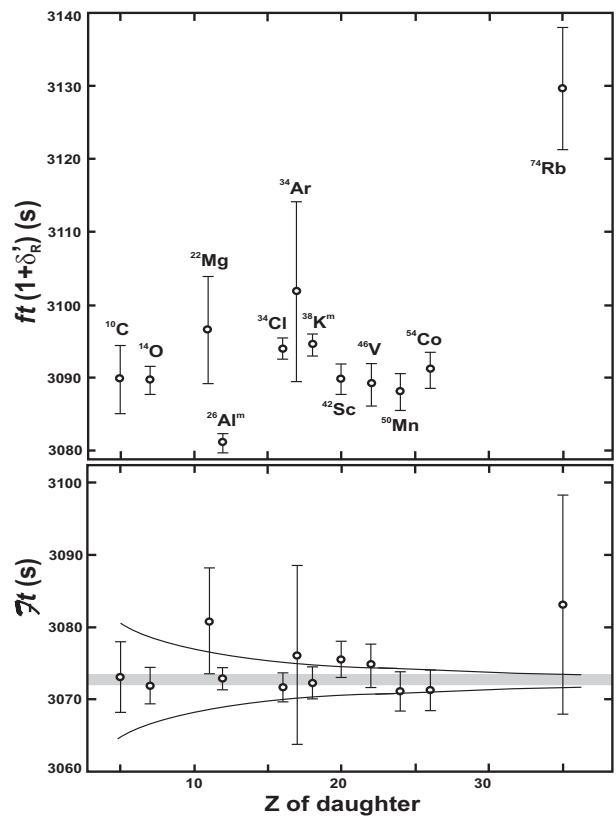,width=8cm}
\caption{In the top panel are plotted the experimental $ft$-values corrected only for $\delta_R^{\prime}$,
those radiative effects that are independent of nuclear structure.  In the bottom panel, the
corresponding $\F t$ values are given; they differ from the top panel simply by inclusion of the
nuclear-structure-dependent corrections, $\delta_{NS}$ and $\delta_C$.  (See Eq.~\ref{Ftconst}.)  Note
that the increased uncertainty for $^{74}$Rb reflects the lack of experimental constraints on the
nuclear shell model in this mass region.   The
horizontal grey band in the bottom panel indicates the average $\overline{\F t}$ value with its
uncertainty.  The curved lines represent the approximate loci the $\F t$ values would follow if an
induced scalar current existed with values of $\fS = \pm 0.002$ in electron rest-mass units or
a fundamental scalar current with $C_S/C_V = \pm 0.002$.}
\label{fig1}
\end{figure}

{\it Nuclear-structure-dependent corrections.}  As can be seen from Eq.~\ref{Ftconst}, the $\F t$ values
differ from their $ft$-value counterparts by the application of the $\delta_R^{\prime}$ radiative correction
and the nuclear-structure-dependent corrections, $\delta_{NS} - \delta_C$.  To isolate and illustrate the
effect of the latter corrections, we plot in the top panel of Fig.~\ref{fig1} the values of $ft(1 +
\delta_R^{\prime})$ for the 12 transtions listed in Table~\ref{table1} and the corresponding values of
$\F t$ in the bottom panel.  They differ only by the inclusion of the structure-dependent corrections in the
latter and, obviously, these corrections act very well to remove the
considerable ``scatter" that is apparent in the top panel and is effectively absent in the bottom one.  It is
important to note that, with the single exception of $^{74}$Rb, the calculations of $\delta_{NS}$ and
$\delta_C$ \cite{To02} are based on well-established shell-model wave functions, and were further tuned to
measured binding energies, charge radii and coefficients of the isobaric multiplet mass equation.  Their
origins are completely independent of the superallowed decay data.  Thus the consistency of the corrected
$\F t$ values appearing in the bottom panel of the figure is a powerful validation of the calculated
corrections used in their derivation.

{\it CVC(1) -- vector coupling constant.}  The CVC hypothesis -- that the weak vector current is just an
isospin rotation of a conserved electromagnetic vector current -- has several direct implications.  The first is
that $\GV$ is indeed constant, independent of the nuclear medium.  The fact that the $\F t$ values, listed in
Table~\ref{table1} and plotted in the bottom panel of Fig.~\ref{fig1}, form a consistent set over a wide
range of nuclei, $10 \leq A \leq 74$, verifies this expectation of CVC at the level of $3 \times 10^{-4}$, 
which is the fractional uncertainty we obtain for $\overline{\F t}$.  This is a 30\% improvement over the
best previous value \cite{Ha90} -- also obtained from superallowed beta decay -- and can be attributed to
improvements in the experimental data.

{\it CVC(2) -- induced scalar current.}  A second implication of the CVC hypothesis is that the ``induced"
scalar term in the vector part of the weak interaction -- as distinct from any possible ``fundamental" scalar
current -- must be zero.  An independent argument \cite{We58}, that there be no second-class currents in the
hadronic weak interaction, also requires it to vanish.  The induced scalar term is characterized by the
coupling constant $\fS$ (using the notation of Behrens and B\"{u}hring \cite{Be82}), the presence of which
would affect the calculation of the statistical-rate function, $f$, {\it via} a term in the shape-correction
function that is inversely proportional to the positron energy.  Since the total superallowed-transition decay energy increases
with $Z$, an induced scalar term would therefore have its greatest effect on the $\F t$ values at low $Z$, 
introducing curvature in that region.  The bottom panel of Fig.~\ref{fig1} shows the loci of $\F t$ values
that would be expected if $\fS = \pm 0.002$.  Obviously, the $\F t$ values do not exhibit any such curvature
and, from a least-square fit to the data, we obtain, $\fS = -0.00005(130)$ or, expressed as a limit, 
$|\fS| \leq 0.0013$, in electron rest-mass units.

{\it Fundamental scalar current.}  Although the weak interaction is normally described by an equal mix of
vector- and axial-vector interactions, which maximizes parity violation, there is no {\it a priori} reason
to rule out other forms of fundamental couplings, notably scalar and tensor interactions.  It turns out that
a ``fundamental" scalar current would have the same effect on the $\F t$ values as would an ``induced" scalar
term: it would introduce curvature in the loci of values at low $Z$ as illustrated in Fig.~\ref{fig1}.  The
superallowed data in this case yield the result $C_S/C_V = -0.00005(130)$, or a limit of $|C_S/C_V|
\leq 0.0013$,  as expressed in the conventional notation of Jackson, Trieiman and Wyld \cite{Ja57}.  The
corresponding result for the Fierz interference constant is $b_F = +0.0001(26)$.  These results represent
a factor of 30 reduction in the central values compared to our previously published result \cite{To03} with
the standard deviation being essentially unchanged.  These are by far the most stringent limits on $|C_S/C_V|$
or $|b_F|$ ever obtained from nuclear beta decay.

{\it The $V_{ud}$ element of the CKM matrix.}  With a mutually consistent set of $\F t$ values, we can now insert
their average value \footnote{For this purpose we use just the nine most accurate $\F t$
values.  Further, we also make a small adjustment \cite{Ha05} to account for
possible systematic uncertainties in the calculated values of $\delta_C$ by
averaging our values given in Table~\ref{table1} with those of Ormand and Brown \cite{OB95} and increasing the error assigned to $\overline{\F t}$.  We use
$\overline{\F t} = 3073.5(12)$.},
$\overline{\F t}$, into Eq.~\ref{Ftconst} and determine the vector coupling constant $\GV$
using the value $\DRV = 2.40(8)\%$ calculated for the transition-independent radiation correction by Marciano
and Sirlin \cite{MS2}.  
%(We also make a small adjustment \cite{Ha05} to account for possible systematic uncertainties in
%the calculated values of $\delta_C$.) 
The value of $\GV$ itself is of little interest but, when combined with $\GF$, the
weak interaction constant for the purely leptonic muon decay, it yields a value for the element $V_{ud}$ of the CKM
matrix: $V_{ud} = \GV/\GF$.  Taking the Particle Data Group (PDG) value \cite{PDG04} of $\GF /(\hbar c )^3 = 1.16639(1)
\times 10^{-5}$ GeV$^{-2}$, we obtain $|V_{ud}| = 0.9738(4)$.  Compared to our previously
recommended value \cite{To03}, this result differs by two units in the last digit quoted and has a reduced
uncertainty.  It also agrees with, but is considerably more precise than, the result obtained using the same
statistical procedures from the world data for neutron decay \cite{To03}: {\it viz.} $|V_{ud}| = 0.9745(16)$.  
Note that, by more than an order of magnitude, $V_{ud}$ is the most precisely determined element of the CKM matrix.

{\it CKM unitarity -- top row.}  The CKM matrix transforms the basis of quark mass eigenstates to that of the quark
weak-interaction eigenstates, and it must therefore be unitary in order that the basis remains orthonormal.  An
important test of the standard model is whether the experimentally determined matrix elements indeed satisfy a
unitarity condition.  Currently, the sum of the squares of the top-row elements, which should equal one, constitutes
the most demanding such test available.  With the value just obtained for $|V_{ud}|$ and the PDG's recommended
values \cite{PDG04} of $|V_{us}| = 0.2200(26)$ and $|V_{ub}| = 0.00367(47)$, this test yields
\be
|V_{ud}|^2 + |V_{us}|^2 + |V_{ub}|^2 = 0.9966 \pm 0.0014,
\label{sumfail}
\ee
which fails unitarity by 2.4 standard deviations.  A recent measurement of the $K^+ \rightarrow \pi^0 e^+ \nu_e ~ (
K_{e3}^+)$ branching ratio from the Brookhaven E865 experiment \cite{Sh03} obtains $V_{us} = 0.2272 \pm 0.0030$.  If
this value alone were adopted for $V_{us}$ rather than the PDG average of many experiments, the sum in Eq.~\ref{sumfail}
would equal 0.9999(16) and unitarity would be fully satisfied.

{\it CKM unitarity -- first column.}  Though for now it is a less demanding test, the first column of the CKM matrix
can also be used to test unitarity.  With our value for $V_{ud}$ and the PDG values for $V_{cd}$ and $V_{td}$, the sum
of squares of these three elements is 0.9985(54), in agreement with unitarity.  The larger uncertainty is entirely due
to $V_{cd}$.

{\it Right-hand currents.}  If parity violation were not maximal, then right-hand currents should be
included in the analysis that yields a value for $V_{ud}$.
Thus, we can set an upper limit on the possible role of
right-hand currents by attributing any apparent non-unitarity of the CKM matrix entirely to that source.  We express the result in terms
of the right-left and left-left coupling constants, $a_{RL}$ and $a_{LL}$, defined by Herczeg \cite{He2}.  If we accept
the unitarity test in Eq.~\ref{sumfail}, then we find $Rea_{LR}/a_{LL} = -0.00176(74)$.  Within the context of the
manifest left-right symmetric model, this result corresponds to a mixing angle of $\zeta = 0.00176(74)$.  If, instead, 
we adopt the E865 value for $V_{us}$, the result becomes $Re a_{LR}/a_{LL} = -0.00007(84)$.

The accumulated world data on superallowed $0^{+} \rightarrow 0^{+}$ $\beta$ decay comprises the results of over one
hundred measurements of comparable precision \cite{Ha05}.  Virtually all the important experimental parameters used as
input to the $ft$-value determinations have been measured in at least two, and often four or five, independent experiments.
Obviously, just another measurement will not have much impact on the precision of the weak-interaction parameters
quoted here.  Nevertheless, it is still possible for well selected experiments with existing or currently foreseen
techniques to make real improvements.  For example, the validation of the nuclear-structure-dependent correction
terms, $\delta_{NS}$ and $\delta_C$, exemplified by the comparison of the two panels in Fig.~\ref{fig1}, can be
improved by the addition of new transitions selected from amongst those with large calculated corrections.  If the
$ft$-values measured for cases with large calculated corrections also turn into corrected $\F t$ values that are
consistent with the others, then this must verify the calculations' reliability for the existing cases, which have
smaller corrections.  In fact, the cases of $^{34}$Ar and $^{74}$Rb, which have only recently been measured, were
chosen for this very reason and, although their precision does not yet equal that of the others, they do indicate that
the corrections so far are living up to expectations.  The precision of these new measurements can certainly be improved, 
and other new cases with large calculated corrections \cite{To02}, such as $^{18}$Ne, $^{30}$S and $^{62}$Ga, would
be valuable additions.

Another area of potential improvement is in the limit set on scalar currents.  The bottom panel of Fig.~\ref{fig1}
clearly illustrates the sensitivity of the low-Z cases to the possible presence of scalar currents.  Reduced
uncertainties, particularly on the decays of $^{10}$C and $^{14}$O, could thus reduce the scalar-current limit
significantly.

Finally, it is important to point out that the biggest contribution to $V_{ud}$ uncertainty -- ~90\% of it, in
fact -- comes from the calculation of the transition-independent part of the radiative correction \cite{MS2}, 
$\DRV = 2.40(8)\%$.  Not only is the uncertainty quoted for $\DRV$ the principal limitation on the precision with
which $V_{ud}$ can be determined from nuclear superallowed $\beta$ decay, but it will have a similar limiting effect
on the determination of $V_{ud}$ from neutron and pion decays as well.  No significant improvement in the precision
of the $V_{ud}$ contribution to the CKM unitarity test can come from any source without a more definitive calculation
of $\DRV$.

\acknowledgments

The work of JCH was supported by the U. S. Dept. of Energy under Grant DE-FG03-93ER40773 and by the Robert A. Welch
Foundation.  IST would like to thank the Cyclotron Institute of Texas A\&M University for its hospitality during
several two-month visits.

\end{document}